\documentstyle[12pt]{article}
\def\({\c c}
\def\|{\'\i }
\def\sqr#1#2{{\vcenter{\hrule height.#2pt
     \hbox{\vrule width.#2pt height#1pt \kern#1pt
      \vrule width.#2pt} \hrule height.#2pt}}}

\begin{document}
\title{Nonholonomic constraints and Voronec's equations}
\author{ Nivaldo A. Lemos\\
{\small{Departamento de F\'{\i}sica}}\\
{\small{Universidade Federal Fluminense}}\\
{\small{Av. Litor\^anea s/n, Boa Viagem }}\\
{\small{24210-340 Niter\'oi, RJ}}\\
{\small{Brazil}}}
\maketitle

\begin{abstract}
Is it allowed, in the context of the Lagrange multiplier formalism, to assume that nonholonomic constraints
are already in effect while setting up  Lagrange's function?
This procedure is successfully applied  in  a recent book  [L. N. Hand and  J. D. Finch,
  {\it Analytical Mechanics}] to the problem of the rolling penny, but it does not work in general,
  as we show by means of a counterexample.
It turns out that in many cases the use of nonholonomic constraints in the process of construction
of the Lagrangian is allowed,
but then the correct equations of motion are  the little known Voronec's equations.

\end{abstract}

\newpage

In the Lagrange
multiplier formalism as applied to nonholonomic systems, the Lagrangian is written as if there were no
constraints. The nonholonomic constraints are taken into account in the formulation of the equations of
motion, but not during the construction of the Lagrangian. Setting up the Lagrangian assuming that
 the constraints are already  in effect is completely equivalent to substituting the  constraint
 equations into the Lagrangian  written as if there were no constraints.
It is  tempting  to take it for granted   that the ensuing reduced Lagrangian
together with the relevant
constraint equations always lead to the correct equations of motion for the system. The procedure just
described  is successfully used in a recent book \cite{Hand} to solve the problem of a penny
rolling  on an inclined plane.
Unfortunately, contrary to what the mentioned book appears to suggest,
this approach   is not valid in general, as we proceed to show with the help of a counterexample.

Consider  a homogeneous sphere rolling without slipping on a horizontal plane. This problem
is treated by the Lagrange multiplier method in \cite{Saletan}. Let $X,Y,Z$ be cartesian axes fixed in
space with the $Z$-axis perpendicular to the plane. The principal moments of inertia with respect to the
center of the sphere are all equal to $2mR^2/5$. With $x,y$ the coordinates of the center of the sphere,
the Lagrangian, being equal to the kinetic energy, is given by

\begin{equation}
\label{lagrangian1}
L=\frac{m}{2}({\dot x}^2 + {\dot y}^2) +\frac{mR^2}{5}{\mbox{\boldmath $\omega$}}^2
\,\,\, .
\end{equation}
The constraint equations are

\begin{equation}
\label{constraints1}
{\dot x}= R \omega_y = R {\dot \theta} \sin\phi - R {\dot \psi} \sin\theta \cos\phi\,\,\, ,\,\,\,
{\dot y}= -R \omega_x = -R {\dot \theta} \cos\phi - R {\dot \psi} \sin\theta \sin\phi
\,\,\, .
\end{equation}
In terms of the Euler angles $\phi,\theta,\psi$ the Lagrangian (\ref{lagrangian1}) takes the form

\begin{equation}
\label{lagrangian2}
L=\frac{m}{2}({\dot x}^2 + {\dot y}^2) +\frac{mR^2}{5}({\dot \phi}^2  + {\dot \theta}^2 +
{\dot \psi}^2+ 2{\dot \phi}{\dot \psi}\cos\theta )
\,\,\, .
\end{equation}

According to the method employed in \cite{Hand}, which takes into account the rolling constraint in
the construction of  the Lagrangian, the kinetic energy is written in terms of the
rotational degrees of freedom alone by taking the moments of inertia with respect to the contact
point of the sphere with the plane. The Lagrangian is now

\begin{equation}
\label{reducedlagrangian1}
{\bar L}=\frac{1}{2}\,\frac{7mR^2}{5}\,(\omega_x^2 + \omega_y^2) +\frac{1}{2}\,\frac{2mR^2}{5}\,\omega_z^2
\,\,\, .
\end{equation}
This is exactly the   Lagrangian that one
obtains by inserting the constraint equations (\ref{constraints1}) into  the Lagrangian (\ref{lagrangian1}),
which was written as if there were no constraints.
In terms of the Euler angles the reduced Lagrangian (\ref{reducedlagrangian1}) becomes

\begin{equation}
\label{reducedlagrangian2}
{\bar L}=\frac{7mR^2}{10}({\dot \theta}^2 + {\dot \psi}^2\sin^2\theta ) +\frac{mR^2}{5} (   {\dot \phi}^2 + {\dot \psi}^2\cos^2\theta + 2 {\dot \phi}{\dot \psi}\cos\theta )\,\,\, .
\end{equation}

Since the variables $x$ and $y$ do not appear in $\bar L$, according to the reasoning in
\cite{Hand} the constraint equations (\ref{constraints1}) are no longer relevant to
the formulation of the equations  of motion. In particular, the Lagrange equation for $\theta$ is

\begin{equation}
\label{equationtheta1}
\frac{d}{dt}\left(\frac{\partial {\bar L}}{\partial {\dot \theta}}\right)
-\frac{\partial {\bar L}}{\partial \theta}=0 \,\,\, \Longrightarrow \,\,\, 7{\ddot \theta} - 5{\dot\psi}^2\sin\theta\cos\theta  + 2{\dot\phi}{\dot\psi}\sin\theta=0
\,\,\, .
\end{equation}

The treatment of this problem by the Lagrange multiplier method shows that  the two Lagrange
multipliers vanish \cite{Saletan}. According to equation (58$d$) of
\cite{Saletan} the correct equation of motion for $\theta$ is

\begin{equation}
\label{equationtheta2}
{\ddot \theta} + {\dot\phi}{\dot\psi}\sin\theta =0
\,\,\, ,
\end{equation}
which is completely different from equation (\ref{equationtheta1}). The differential equations
(\ref{equationtheta1}) and (\ref{equationtheta2}) generally yield different solutions for $\theta$
because  $\phi,\theta,\psi,{\dot \phi},{\dot \theta},{\dot \psi}$ can be arbitrarily chosen at any
particular instant $t_0$. It is clear, therefore, that the approach
suggested in \cite{Hand} lacks generality, since it works for the  rolling penny but
fails for the  rolling sphere.

In a previous paper \cite{Lemos} we remarked that it is possible to perform a reduction of the
 Lagrangian  taking into account the constraints, but in this case the correct equations of
 motion are Voronec's equations. Given a dynamical system described by the configuration variables
$q_1,\ldots,q_n$, suppose the first $m$ velocities are independent and the $k=n-m$ remaining velocities
can be expressed in terms of the independent ones by means of the equations

\begin{equation}
\label{constraints2}
{\dot q}_{m+l}-\sum_{j=1}^{m} a_{lj} {\dot q}_j = 0\,\,\, , \,\,\, l=1,\ldots,k\,\,\, ,
\end{equation}
where the coefficients $a_{lj}$ are functions of the generalized coordinates $q_1,\dots,q_n$.

Let $L$ be the Lagrangian written without taking
into account the nonholonomic constraint equations (\ref{constraints2}). If the last  $k$  velocities
are eliminated from the Lagrangian by means of equations (\ref{constraints2}), a reduced Lagrangian
${\bar L}$
results:

\begin{equation}
\label{reducedlagrangian}
L(q_1,\dots ,q_n,{\dot q}_1,\dots,{\dot q}_n,t) = {\bar L}(q_1,\dots,q_n,{\dot q}_1,\dots ,{\dot q}_m,t)
\,\,\, .
\end{equation}
Voronec's equations of motion are \cite{Lemos,Fufaev}

\begin{equation}
\label{voronecequations}
\frac{d}{dt}\left(\frac{\partial {\bar L}}{\partial {\dot q}_i}\right)
-\frac{\partial {\bar L}}{\partial q_i}=
\sum_{\nu =1}^{k}\frac{\partial {\bar L}}{\partial q_{m+\nu}} a_{\nu i} +
\sum_{\nu =1}^{k}\sum_{j =1}^{m}\frac{\partial L}{\partial {\dot q}_{m+\nu}}\,b_{ij}^{\nu}\,{\dot q}_j
\,\,\, ,\,\,\, i=1,\ldots,m\,\,\, ,
\end{equation}
where

\begin{equation}
\label{bnuij}
b_{ij}^{\nu}= \frac{\partial a_{\nu i}}{\partial q_j}- \frac{\partial a_{\nu j}}{\partial q_i}
+\sum_{\mu =1}^{k}\left( \frac{\partial a_{\nu i}}{\partial q_{m+\mu}}a_{\mu j}
-\frac{\partial a_{\nu j}}{\partial q_{m+\mu}}a_{\mu i}\right)\,\,\, .
\end{equation}

Setting $q_1=\phi, q_2=\theta, q_3=\psi, q_4=x, q_5=y$, in the present case $m=3$ and $k=2$. The
constraint equations (\ref{constraints1}) can be written in the
form (\ref{constraints2}) with

\begin{equation}
\label{anuijsphere}
a_{11}=0\,, \, a_{12}=R\sin\phi\, , \, a_{13}=-R \sin\theta\cos\phi \, , \,
a_{21}=0\, , \, a_{22}=-R\cos\phi \, , \,a_{23}=-R \sin\theta\sin\phi \,\,\, .
\end{equation}
The definition (\ref{bnuij}) furnishes immediately the only nonvanishing coefficients
$b_{ij}^{\nu}$:

\begin{equation}
\label{bnuijsphere1}
b_{12}^{1}= -b_{21}^{1}= -R\cos\phi \,\,\, , \,\,\, b_{13}^{1}= -b_{31}^{1}= -R\sin\theta\sin\phi
\,\,\, , \,\,\, b_{23}^{1}= -b_{32}^{1}= R\cos\theta\cos\phi\,\,\, ,
\end{equation}
\begin{equation}
\label{bnuijsphere2}
b_{12}^{2}= -b_{21}^{2}= -R\sin\phi \,\,\, , \,\,\, b_{13}^{2}= -b_{31}^{2}= R\sin\theta\cos\phi
\,\,\, , \,\,\, b_{23}^{2}= -b_{32}^{2}= R\cos\theta\sin\phi\,\,\, .
\end{equation}

It follows that Voronec's equation for $\theta$ is

\begin{equation}
\label{voronectheta1}
\frac{7mR^2}{5}{\ddot \theta} - mR^2{\dot\psi}^2\sin\theta\cos\theta  + \frac{2mR^2}{5}{\dot\phi}{\dot\psi}\sin\theta
=m{\dot x}(b_{21}^1{\dot\phi} + b_{23}^1{\dot\psi})+
m{\dot y}(b_{21}^2{\dot\phi} + b_{23}^2{\dot\psi})
\,\,\, .
\end{equation}
The use of the constraint equations (\ref{constraints1}) and a little algebra reduce the above  equation to

\begin{equation}
\label{voronectheta2}
{\ddot \theta} + {\dot\phi}{\dot\psi}\sin\theta =0
\,\,\, ,
\end{equation}
which coincides with the correct equation for $\theta$ furnished by the Lagrange multiplier method.
It is equally straightforward to check that the remaining Voronec equations for $\phi$ and $\psi$
coincide with those  obtained by the Lagrange multiplier method.

In short, the method employed in \cite{Hand} is not valid in general and should not be taught to students.
The correct result obtained for the rolling penny is the product of a mere accident.

As a general rule, the use of nonholonomic constraints while
setting up the Lagrangian is allowed, but the correct equations of motion are Voronec's equations,
and not the ones given by the Lagrange multiplier method.

\end{document}